\documentclass[twocolumn]{aastex701}

\PassOptionsToPackage{numbers, compress}{natbib}
\newcommand{\model}{\texttt{Minuet}}

\usepackage[utf8]{inputenc} 
\usepackage[T1]{fontenc}    
\usepackage{hyperref}       
\usepackage{amsmath}
\usepackage{url}            
\usepackage{booktabs}       
\usepackage{amsfonts}       
\usepackage{nicefrac}       
\usepackage{microtype}      
\usepackage{graphicx}
\usepackage{xcolor}         

\begin{document}

\title{\model{}: A Diffusion Autoencoder for Compact Semantic Compression of Multi-Band Galaxy Images}

\author[0000-0003-4906-8447]{Alexander~T.~Gagliano}
\affiliation{The NSF AI Institute for Artificial Intelligence and Fundamental Interactions}
\email{gaglian2@mit.edu}
\affiliation{Center for Astrophysics \textbar{} Harvard \& Smithsonian, 60 Garden Street Cambridge, MA 02138 \\}
\affiliation{Department of Physics and Kavli Institute for Astrophysics and Space Research, Massachusetts Institute of Technology, 77 Massachusetts Avenue, Cambridge, MA 02139, USA}

\author[0000-0003-2779-6507]{Yunyi~Shen}
\email{yshen99@mit.edu}
\affiliation{Department of Electrical Engineering and Computer Science, Massachusetts Institute of Technology, 77 Massachusetts Ave. Cambridge, MA 02139}
\affiliation{The NSF AI Institute for Artificial Intelligence and Fundamental Interactions}

\author[0000-0002-5814-4061]{V.~A.~Villar}
\email{ashleyvillar@cfa.harvard.edu}
\affiliation{The NSF AI Institute for Artificial Intelligence and Fundamental Interactions}
\affiliation{Center for Astrophysics \textbar{} Harvard \& Smithsonian, 60 Garden Street Cambridge, MA 02138 \\}

\begin{abstract}
 The Vera C. Rubin Observatory is slated to observe nearly 20 billion galaxies during its decade-long Legacy Survey of Space and Time. The rich imaging data it collects will be an invaluable resource for probing galaxy evolution across cosmic time, characterizing the host galaxies of transient phenomena, and identifying novel populations of anomalous systems. While machine learning models have shown promise for extracting galaxy features from multi-band astronomical imaging, the large dimensionality of the learned latent space presents a challenge for mechanistic interpretability studies. In this work, we present \model{}, a low-dimensional diffusion autoencoder for multi-band galaxy imaging. \model{} is trained to reconstruct 72x72-pixel $grz$ image cutouts of 6M galaxies within $z<1$ from the Dark Energy Camera Legacy Survey using only five latent dimensions. By using a diffusion model conditioned on the transformer-based autoencoder's output for image reconstruction, we achieve semantically-meaningful latent representations of galaxy images while still allowing for high-fidelity, probabilistic reconstructions. We train a series of binary classifiers on \model{}'s latent features to quantify their connection to morphological labels from Galaxy Zoo, and a conditional flow to produce posterior distributions of SED-derived redshifts, stellar masses, and star-formation rates. We further show the value of \model{} for nearest neighbor searches in the learned latent space. \model{} provides strong evidence for the low intrinsic dimensionality of galaxy imaging, and introduces a class of astrophysical models that produce highly compact representations for diverse science goals.
\end{abstract}

\section{Introduction}
The chemical and dynamical histories of galaxies sit at the nexus of diverse astrophysical domains. Optical observations of galaxies have been critical for probing dust formation \citep{2000Calzetti_Dust}, quenching \citep{2010Peng_Quenching}, and baryonic feedback across cosmic time \citep{2008Baldry_Baryon,2010Genzel_z1Feedback}. A galaxy's star-formation history also strongly impacts the nature of the transient phenomena that it can host \citep{2006Fruchter_GRBs,2016French_TDEs,2016Galbany_HII,2021Schulze_PTF}, and host-galaxy correlations have been used to constrain the progenitor systems of core-collapse SNe \citep{2021Schulze_CCSNe,2025Ganss_SESNe} and peculiar SNe~Ia \citep{2019Panther_91bg}. Galaxy clusters have also been used to place constraints on cosmological parameters \citep{2017Alam_SDSSCosmo}.

The broad-band imaging data obtained by wide-field surveys such as the Sloan Digital Sky Survey \citep{2025SDSS} and Pan-STARRS \citep{2016Chambers_PanSTARRS}, have been an essential contribution to diverse galaxy studies. Population-level studies of these datasets have consistently uncovered unexpected populations that upended existing evolutionary theory (e.g., ``little red dots'' uncovered by \textit{JWST}, \citealt{2024Matthee_LRDs}; and ``green peas'' and ``red spirals'' identified by citizen scientists through Galaxy Zoo\footnote{\url{https://www.zooniverse.org/projects/zookeeper/galaxy-zoo/}}; \citealt{2009Cardamone_GreenPeas,2010Masters_RedSpirals}).

Galaxy imaging has also provided a rich test bed for developing and adapting machine learning architectures for astrophysics. There exists a well-established history of convolutional neural networks trained to classify galaxies morphologically \citep{2021Cavanagh_CNNs}. Techniques for photometric redshift estimation, initially trained on extracted photometry \citep{2014SOMs, 2023Sun_NormalizingFlows,2024Crenshaw_NFs}, have begun to focus instead on multi-band imaging of a target \citep{2025Engel_MantisShrimp}. Deep learning models have been more recently used to synthesize realistic multi-band galaxy images \citep{2025Fan_ScoreBased}, deblend sources \citep{2022Hemmati}, and learn an empirical mapping between survey-specific datasets \citep{2025Ruan_Translation}. 

In parallel to these domain-specific approaches, there has been sustained interest in the machine learning community in exploring model architectures and training objectives that can extract semantically-meaningful, generalizable features from raw imaging data. This philosophy has been inspired by the successes of foundation models in computer vision and natural language processing \citep{2021Bommasani_FMs}. A suite of large astronomical models trained using self-supervised objectives, with one or more input modalities, have shown promising initial results in this direction \citep{2024Parker_AstroCLIP,2024Zhang_FM,2025Audenaert_FMs,2025Euclid_FMs,2025Roy_FMs,2025Zhao_FMs,2025Zuo_FALCO}.

Though the scale of astronomical models continues to increase, the resulting high-dimensional latent spaces often obscure physical interpretation, making it difficult to connect learned features to astrophysically meaningful quantities such as stellar mass, metallicity, or morphology. Unconstrained latent spaces may also encode spurious data features \citep{2000Tishby_Bottlenecks} that degrade performance when applied to data with distributional shifts, especially when trained on observational data subject to complex selection biases and poorly characterized systematics. In the case of reconstruction-based objectives, models encode pixel-level information, which is highly redundant \citep{2021Pope} and does not contain significant semantic content (in contrast with natural language; \citealt{2021He_MaskedAEs}).

In this work, we present \model{}, a model trained to extract a highly compact set of generalizable features from multi-band galaxy imaging. Our model is trained to compress $grz$ images into five latent dimensions using a self-supervised objective. The architecture of \model{} is inspired by \citet{2021Preechakul_DiffAEs} and employs a transformer encoder and a conditional probabilistic diffusion decoder. The resulting model is able to simultaneously extract semantically-meaningful features without supervised training and still achieve accurate galaxy reconstructions.

Our paper is structured as follows. In \textsection\ref{sec:data}, we describe the datasets used to train the model. In \textsection\ref{sec:architecture}, we expand on \model{}'s architecture and our training setup. We investigate correlations between the learned latent space and both user-labeled morphological and SED-inferred physical properties of galaxies in our validation set, and train a series of lightweight models to infer these properties from the encoded features. We conclude by discussing the potential uses of the model and directions for future work in \textsection\ref{sec:discussion}. 

All code used to train and validate \model{}, as well as an interactive visualization of the learned latent space, can be found at the Github repository for this work\footnote{\url{https://github.com/alexandergagliano/Minuet}}.

\section{Data}\label{sec:data}
We select galaxies from the Dark Energy Camera Legacy Survey \citep[DECaLS;][]{2019Dey_DECaLS}. DECaLS reaches a 5$\sigma$ depth of $g\approx24$, making it a sufficient analog for upcoming Rubin data ($g\approx24.5$)\footnote{\url{https://www.lsst.org/scientists/keynumbers}}. To obtain homogeneous imaging data for each galaxy, we retrieve $grz$-band images from the catalog constructed by \citet{2022Stein_Mining}\footnote{\url{https://github.com/georgestein/ssl-legacysurvey}}. We also impose an upper limit on the redshift and the apparent magnitude of each source, to ensure that each galaxy is sufficiently resolved. We consider only galaxies with an apparent $z$-band magnitude of $z<20$, and within $z<1.0$. This redshift range is well-matched to the redshift range of expected SNe in LSST. The resulting sample has a maximum $r$-band magnitude of $r\approx23.0$.

Next, we cross-match the remaining sample to the catalog of physical properties constructed by \citet{2022Zou_SED}. This catalog reports the 50th percentile and the inter-quartile range of the posteriors for redshift, stellar mass, and star-formation rate of each galaxy obtained with the spectral-energy-distribution (SED)-fitting codes \texttt{CIGALE} \citep{2019Boquien_CIGALE}.  Our complete cross-matched sample contains 6,050,000 galaxies. We split the dataset into fractions of 60/20/20 for train/test/validation sets, respectively. 

\begin{figure}
    \centering
    \includegraphics[width=\linewidth]{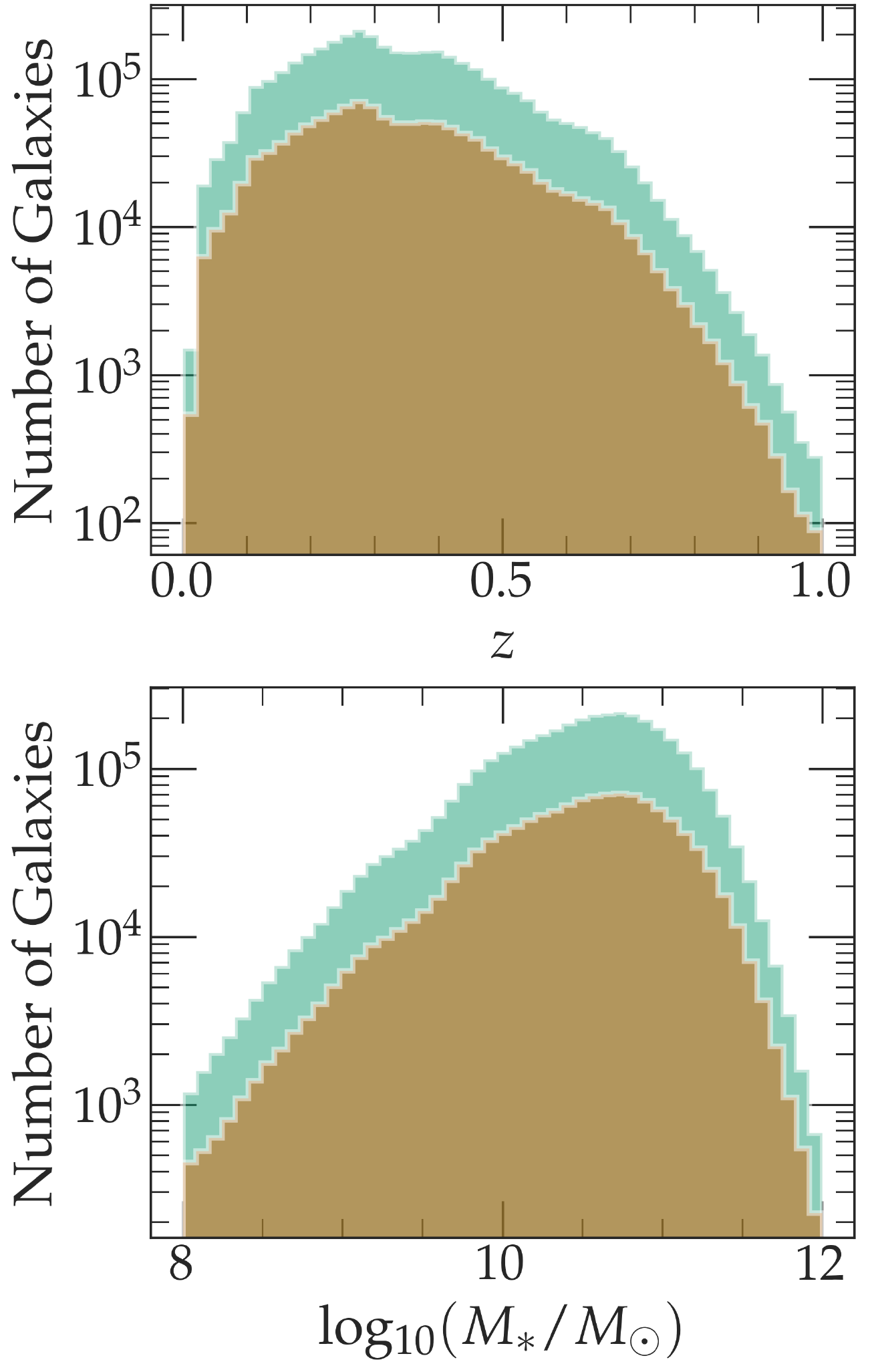}
    \caption{Distribution of redshift (top) and stellar mass (bottom) for galaxies in the training (green) and testing (orange) sets. The validation set is identical in size and distribution to the test set.}
    \label{fig:data_distribution}
\end{figure}

We visualize the distribution of redshifts and stellar mass between the train and test sets after cross-matching in Figure~\ref{fig:data_distribution}.

The \citet{2022Stein_Mining} catalog contains 128x128 pixel postage stamps centered on each DECaLS-detected galaxy in $grz$ filters. We crop each image to the central 72x72 px to reduce training time and reduce the relative contribution of neighboring sources to the score. 

\section{Model Architecture and Training}\label{sec:architecture}

Our objective is to extract features correlated with the physical properties of the galaxy (e.g., the stellar mass, star-formation rate, and redshift) and insensitive to instrumental systematics. We achieve this goal by re-casting our training objective as a conditional generation task. In other words, we aim to reconstruct the galaxy image given a small number of learned latent features. 

\begin{figure*}
    \centering
    \includegraphics[width=\linewidth]{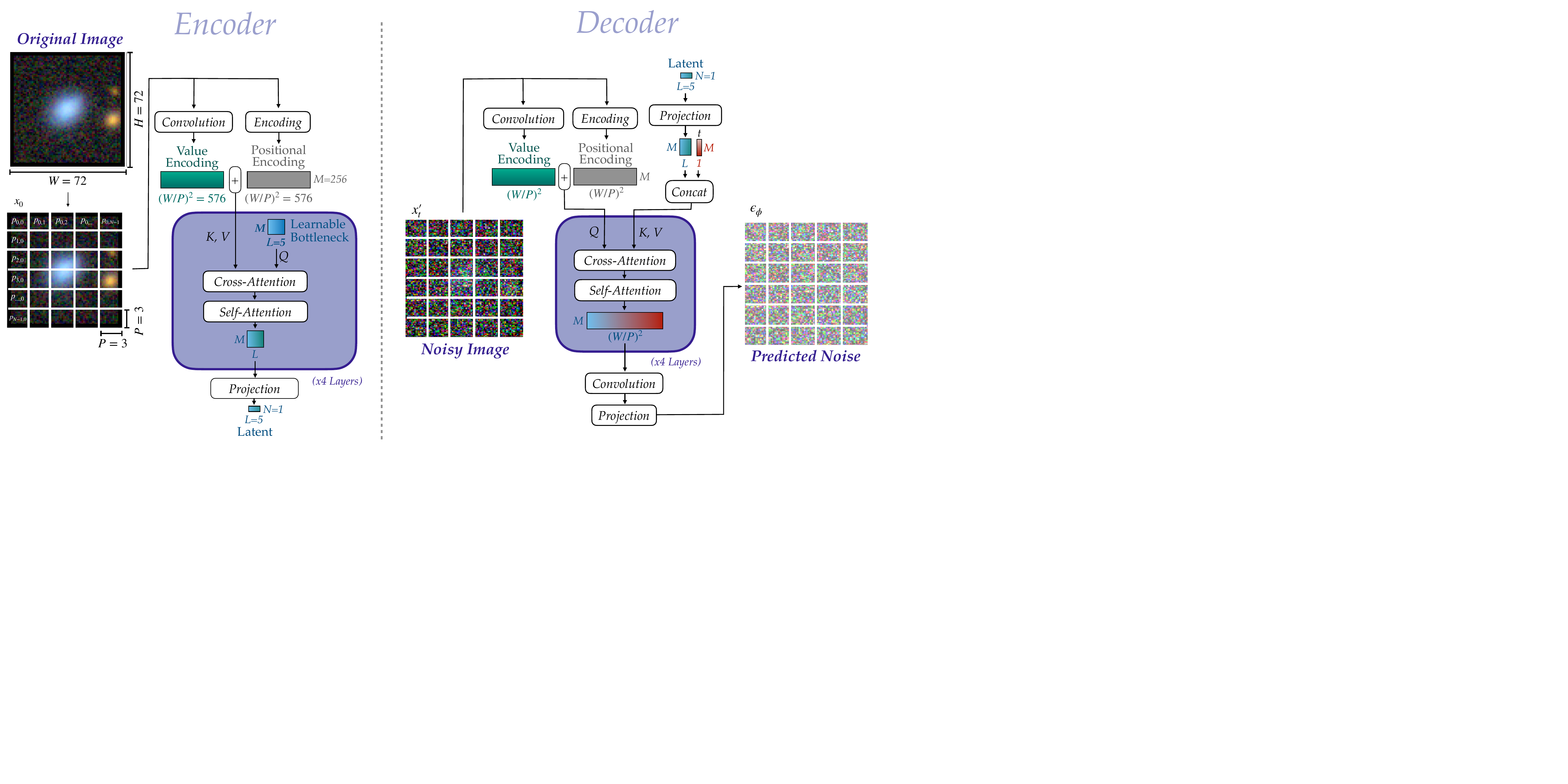}
    \caption{The \model{} architecture. The encoder consists of four Perceiver encoder layers, in which the input tokens as key and value attend to a bottlenecked representation of the input as query. Image inputs have dimensionality $B \times C \times H \times W$, where $C=3$ is the channel dimension, $B=128$ is the batch size, $W=72$ is the image width and $H=72$ is the image height. The intermediate output is a tensor of dimensionality $L\times M$, which is projected to a final output of $L\times1$ latent tokens. In the decoder stage, the latent tokens are concatenated to a learnable embedding of the diffusion time and passed through the same Perceiver modules as the encoder, before finally being projected to the input dimensionality.}
    \label{fig:arch}
\end{figure*}

In this work, we use the Diffusion AutoEncoder-Perceiver (\texttt{daep}) architecture described in \citet{2025Shen_Perceiver} and inspired by recent advancements in computer vision \citep{2021Preechakul_DiffAEs}. \texttt{daep} combines Perceiver layers \citep[a transformer variant,][]{2021Jaegle_Perceiver} with a diffusion-based decoder \citep{2020Song_Diffusion} to extract a semantically-meaningful latent space without sacrificing the reconstruction quality of the decoder.

We provide a schematic of our architecture in Figure~\ref{fig:arch}. The model is parameterized by a model dimension $M$ and a sequence length $L$, and processes $H\times W$ galaxy images with channel dimension $C=3$ (corresponding to $grz$ filters). To tokenize the inputs, each galaxy image is segmented into $(W/P)^2=576$ patches of $3 \times 3$ pixels, and patch values are processed by a convolutional neural network layer to the model dimension and then added to a learnable encoding corresponding to patch positions. These tokens serve as input to four sequential Perceiver encoder layers, which computes a cross-attention matrix between the input tokens and a learnable bottleneck of dimensionality $M \times L$. The final output is projected to a latent representation $z$ of shape $N \times L$ ($N=1$-dimensional features for $L=5$ tokens).

The decoder is a conditional Denoising Diffusion Probabilistic Model \citep[DDPM;][]{2020Ho_ddpm} using four Perceiver decoding layers. The latent tokens are projected back to the model dimension $M$, and concatenated with an encoding of the diffusion time $t$. Cross-attention is computed between noisy image tokens and this latent context (conditioning the diffusion trajectory), then convolution and projection layers convert the output back to the original image dimensionality $H\times W \times C$ to predict the pixel-wise noise added to the original image $\epsilon_t$. A typical diffusion model is trained to minimize the mean-squared error (MSE) between the predicted noise  $\epsilon_{\theta}$ and the true noise $\epsilon_t$ added to the input image $x_0$:  $\|\epsilon_{\theta}(x_t, t, z) - \epsilon_t\|_2^2$. In our case, we adapt this loss to weight each pixel-wise MSE by $\alpha^{i,j,c}$, such that our modified objective is   $$\sum_{i,j,c} \alpha^{i,j,c} \cdot |\epsilon^{i,j,c}_{\theta}(x_t, t, z) - \epsilon^{i,j,c}_{t}|^2$$ 

This coefficient prioritizes the reconstruction of pixels associated with the central galaxy in each image, and downweights the reconstruction of neighboring stars and galaxies. 

We use the \texttt{segmentation} module within the \texttt{Photutils} library to detect sources in each $r$-band image to $3$-sigma and containing a minimum of 10 image pixels. We then deblend sources detected in the images and select the source closest to the image center. Our model's objective is the score of the diffusion decoder, with  $\alpha$ determining the weight of each pixel. We set $\alpha^{i,j,c}=0.9$ for pixels associated with the central galaxy, and $\alpha^{i,j,c}=0.1$ for all other image pixels. 

At inference time, we sample from the learned posteriors using a Denoising Diffusion \textit{Implicit} Model \citep{2020Song_ddim} rather than the stochastic DDPM used in training. DDIM uses the same trained neural network but reformulates the reverse generative process as a deterministic trajectory, enabling us to reduce sampling from 1000 diffusion steps to 50 inference steps while maintaining reconstruction quality. This deterministic sampling ensures the same latent features $z$ always produce identical reconstructions when starting from the same initial noise, enabling direct visual comparison and systematic evaluation. For the stochastic reconstruction analysis in \textsection\ref{subsec:reconstruction_samples}, we draw multiple samples using different random noise realizations.

We list the hyperparameters of our network in Table~\ref{tab:hyp}.

\begin{table}[]
    \centering
    \begin{tabular}{c|c}
       Parameter  & Value \\ \hline
       Bottleneck Length & 5 \\
       Bottleneck Dim. & 1 \\
       Model Dim.  & 256 \\ 
       Latent Dim. & 5 \\
       Patch Size & 3 \\
       $N_{\mathrm{Encoder}}$ & 4 \\
       $N_{\mathrm{Decoder}}$ & 4\\
       Batch Size & 128 \\ 
    \end{tabular}
    \caption{Hyperparameters of the trained network.}
    \label{tab:hyp}
\end{table}

We correct each image for Galactic Reddening using the values provided in the DECaLS DR9 release \citep{2021Schlegel_DR9}. During training, we augment our galaxy images using the modules developed in \citet{2022Stein_Mining}. We redden each image by a random value of $E(B-V)$ drawn from the lognormal fit to the data distribution in \citet{2022Stein_Mining}, and randomly rotate each image\footnote{We explored an additional edge enhance augmentation using \href{https://scikit-image.org/docs/0.25.x/auto_examples/filters/plot_unsharp_mask.html}{unsharp masking}, but found it to be too sensitive to statistical fluctuations in images of galaxies at high redshift.}.

We train for 1000 epochs with early stopping using the standard \verb|Adam| optimizer \citep{2014AdamOptimizer} with a learning rate of $5\times10^{-4}$ on four Nvidia A100 80GB GPUs, confirming convergence of the loss and saving the model weights at the epoch where the validation loss is minimized.

\section{Results}\label{sec:results}
\subsection{Image Reconstruction Performance}\label{subsec:reconstructions}
\begin{figure*}
    \centering \includegraphics[width=\linewidth]{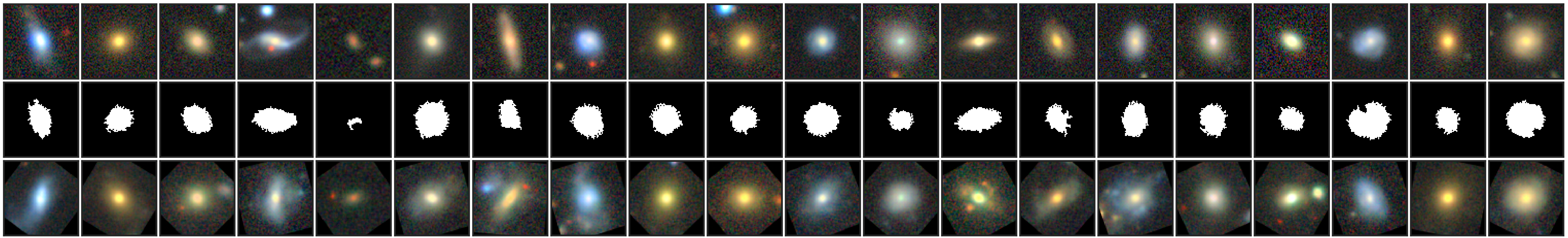}
    \caption{20 DECaLS $grz$-band images of galaxies in our bright ($r<18$) galaxy sub-sample (top row), segmentation masks (middle row), and sample reconstructions from \model{} using 5 latent features (bottom row). Key galaxy color and brightness profile features are preserved, although some high-frequency information is lost and the reconstructions do not preserve galaxy orientation.}
    \label{fig:sample_reconstructions}
\end{figure*}

We present a randomly-selected sample of 20 galaxies within $r<18$ from the validation set in Figure~\ref{fig:sample_reconstructions}, along with their calculated segmentation masks and a single image reconstructed by the model using ddim with a noise schedule featuring 1000 steps. Key color and morphological properties of each galaxy are visually reconstructed, although the model struggles to reconstruct the outer periphery of extended galaxies whose edges have relatively low surface brightness and as a result are not captured by the segmentation mask. We also observe some loss of high-frequency galaxy structure within the mask,  unsurprising given the low-dimensional latent space.

Interestingly, the galaxy samples drawn from the learned posteriors do not capture the orientation of the galaxies in the validation set. Multiple samples drawn from the posterior of a single inclined galaxy (Figure~\ref{fig:ddim_samples}) confirm that the learned latent features are agnostic to galaxy orientation. It is possible that this is due to the limited model capacity, with the galaxy's light profile dominating the gradient on the score function.

\subsection{Variance of Galaxy Reconstructions}\label{subsec:reconstruction_samples}

\begin{figure*}
    \centering
    \includegraphics[width=\linewidth]{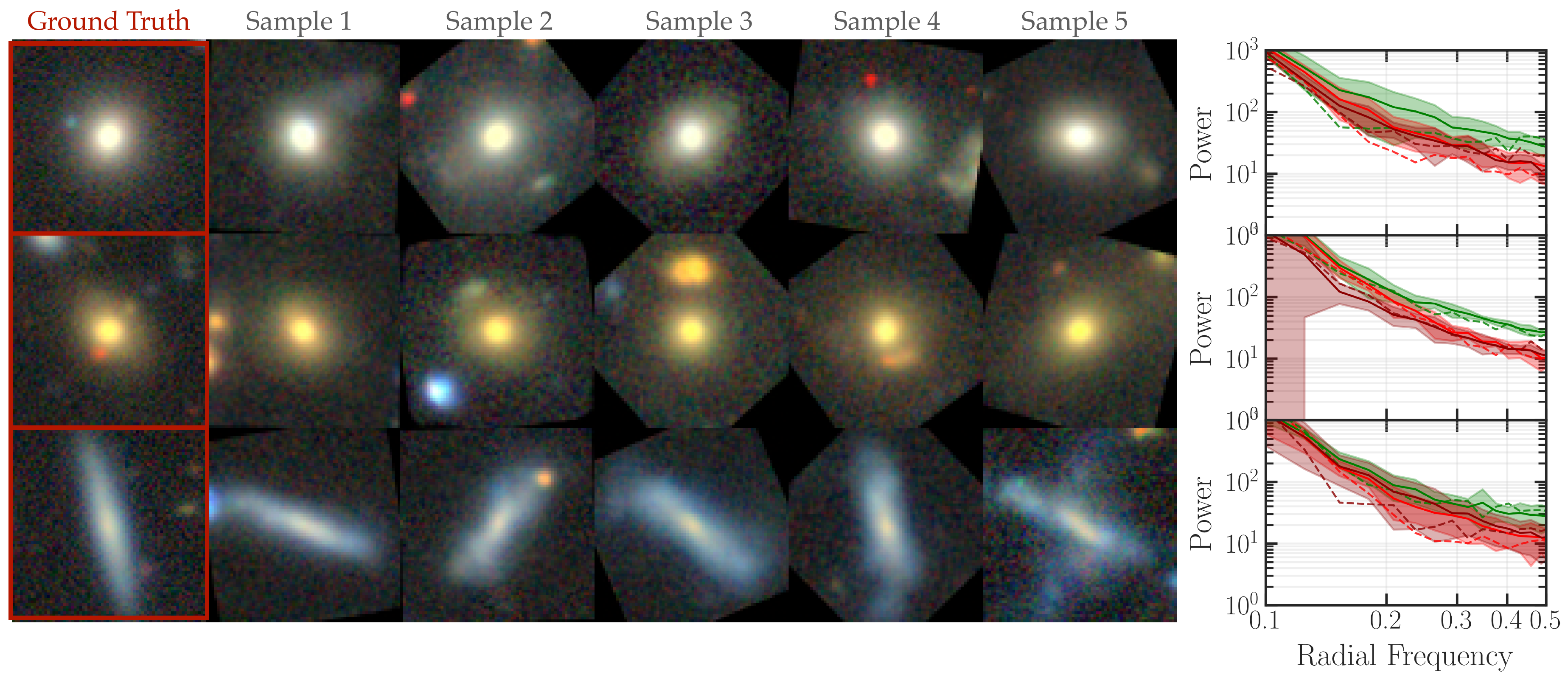}
    \caption{\textbf{Left Column:} Ground truth images of three galaxies from our bright test set. \textbf{Middle  Columns:} Five samples drawn from the learned posterior using DDIM starting from random Gaussian noise realizations (middle five columns). Image samples are agnostic to galaxy orientation (as seen in middle and bottom rows) and neighboring objects due to the segmentation mask applied in training.\textbf{Right Column:} Radial frequency distributions in $grz$ (green, red, and dark red, respectively). Ground truth distributions are shown as dashed lines, and mean and shaded regions correspond to the 1$\sigma$ standard deviation across the five reconstructions.}
    \label{fig:ddim_samples}
\end{figure*}

To evaluate the robustness of our reconstructions, we draw multiple samples from our ddim decoder for the same latent features associated with three randomly-selected galaxies in the validation set. We present the results in Figure~\ref{fig:ddim_samples}. The general properties of each galaxy are preserved across each realization, while the orientation and the presence of neighboring sources (e.g., a bright star in the first sample of the galaxy in the top row) are not. In addition, the low weighting of neighboring pixels still ensures a high-quality reconstruction even where the segmentation mask fails (top row).

Next, we evaluate the quality of the galaxy reconstructions sampled by our learned posteriors. Multiple reconstruction metrics exist in the literature, with the structural similarity index measure \citep{2004Wang_SSIM} being a popular choice due to its robustness to pixel-level variations (in contrast with, e.g., the mean-squared error between true and reconstructed pixel intensity values). Because our reconstructed samples are agnostic to orientation, we instead compare the radially-averaged power spectrum of reconstructed and ground truth images.

Let $I \in \mathbb{R}^{N_y \times N_x}$ be an $N_y\times N_x$ galaxy image with pixel coordinates $(y,x) \in \{0,\ldots,N_y-1\} \times \{0,\ldots,N_x-1\}$, and spatial-frequency indices given by $(v,u)$ on the same index sets. The discrete Fourier transform can then be calculated as (equation 2-61 of \citealt{GonzalezWoods2018}): \begin{align}
F(v,u)
&= \sum_{y=0}^{N_y-1}\sum_{x=0}^{N_x-1}
I(y,x)\, e^{\!-2\pi i\Big(\frac{ux}{N_x}+\frac{vy}{N_y}\Big)},
\label{eq:dft}
\end{align}
and the power spectrum is calculated as
\begin{align}
P(v,u) &= |F(v,u)|^2. \label{eq:power2d}
\end{align}

We average the power calculated for pixels at constant distance from the image center in spatial-frequency space. Letting $c_x=\lfloor N_x/2\rfloor$ and $c_y=\lfloor N_y/2\rfloor$ denote the center indices of the image (where $\lfloor\rfloor$ is the floor function), the discrete radial distance (in pixels) is measured as:

\begin{align}
\rho(v,u) &= \sqrt{(u-c_x)^2 + (v-c_y)^2}. \label{eq:rho}
\end{align}

We set $r_{\max}=\min\{c_x,c_y\}$ and choose $n$ radial bins with edges $0=r_0<r_1<\cdots<r_n=r_{\max}$. In the implementation below we take
$n=r_{\max}$, i.e., each annulus bin is one pixel larger than the previous one, so that $r_i=i$. For each bin $[r_i,r_{i+1})$, we then define the index set of considered pixels as

\begin{align}
\mathcal{A}_i &= \big\{(v,u):\ r_i \le \rho(v,u) < r_{i+1}\big\}.
\end{align}

Finally, the radially-averaged power spectrum is the mean of $P$ over each annulus set $\mathcal{A}$:

\begin{align}
S_i &= \frac{1}{|\mathcal{A}_i|}
\sum_{(v,u)\in \mathcal{A}_i} P(v,u), \qquad i=0,\ldots,n-1.
\label{eq:radavg}
\end{align}

We report each bin at the midpoint radius, and normalize all bins to a minimum value of zero and a maximum of unity.

For each of the three validation-set galaxies in Figure~\ref{fig:ddim_samples}, we  compute the radially-averaged power spectrum of the $grz$-band images. We do the same for each of the five reconstructed images, and compare the mean and 1-$\sigma$ standard deviation across the samples to the ground truth spectrum in the right panel of Figure~\ref{fig:ddim_samples}.

Our power spectra suggest only minor smoothing of our reconstructed images: we find marginally more high-frequency information contained within our ground truth images, and less low-frequency information, relative to our reconstructions. The deviation is strongest for the galaxy in the bottom row of Figure~\ref{fig:ddim_samples}, whose reconstructed halo is smeared relative to the ground truth. This may be a consequence of the segmentation mask used in training, with some low-frequency information leaking into surrounding pixels without significant impact to the batch-averaged score.

\subsection{Inference of Physical Properties}\label{subsec:physical_properties}
\begin{figure*}
    \centering
    \includegraphics[width=\linewidth]{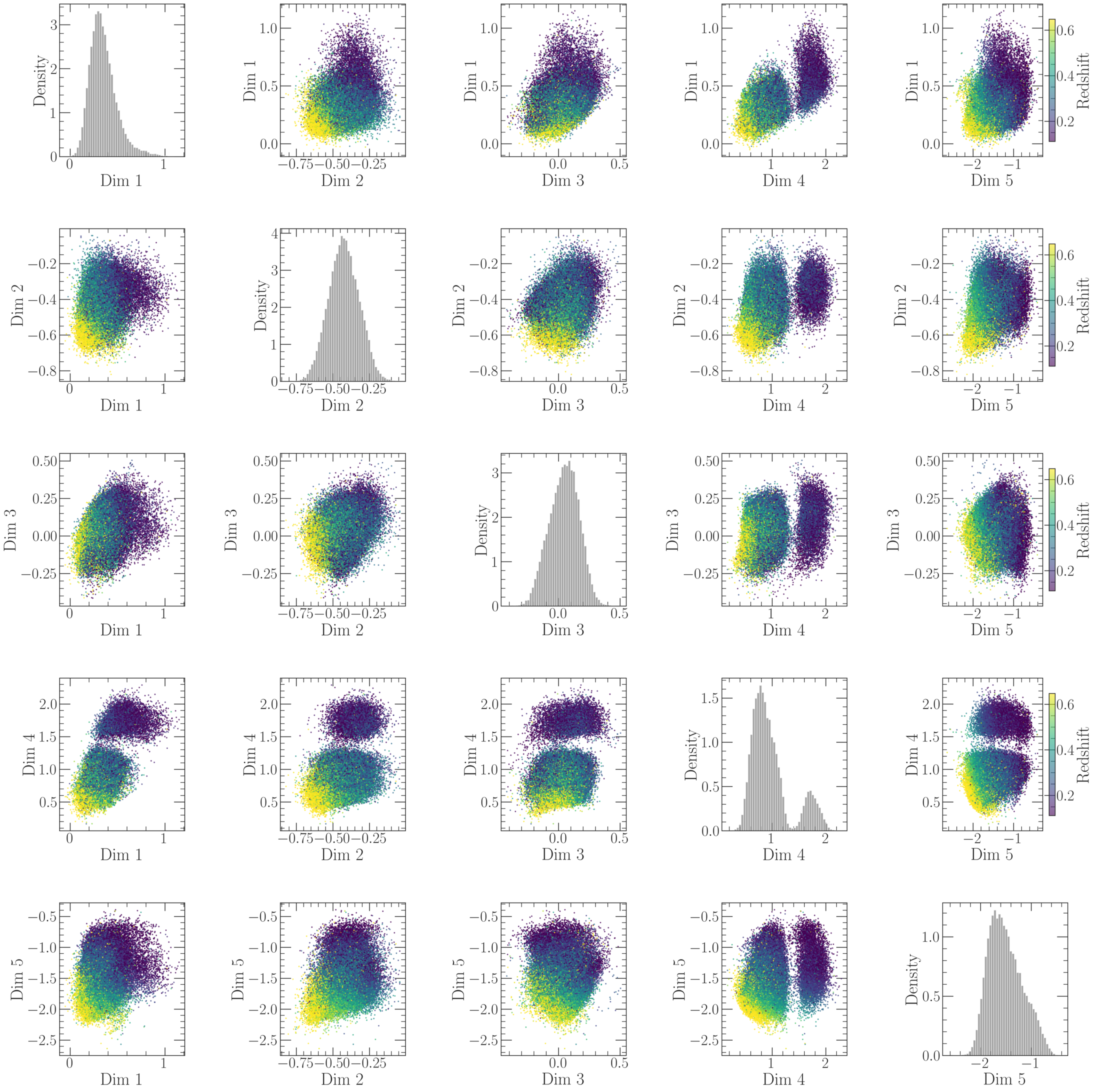}
    \caption{Corner plot of all latent features for the sample of galaxies cross-matched with the catalog from \citet{2022Zou_SED}, and colored by the best redshift for the galaxy (spectroscopic if available, else the photometric value inferred in \citealt{2022Zou_SED} is used).}
    \label{fig:corner_redshift}
\end{figure*}

\begin{figure*}
    \centering
    \includegraphics[width=\linewidth]{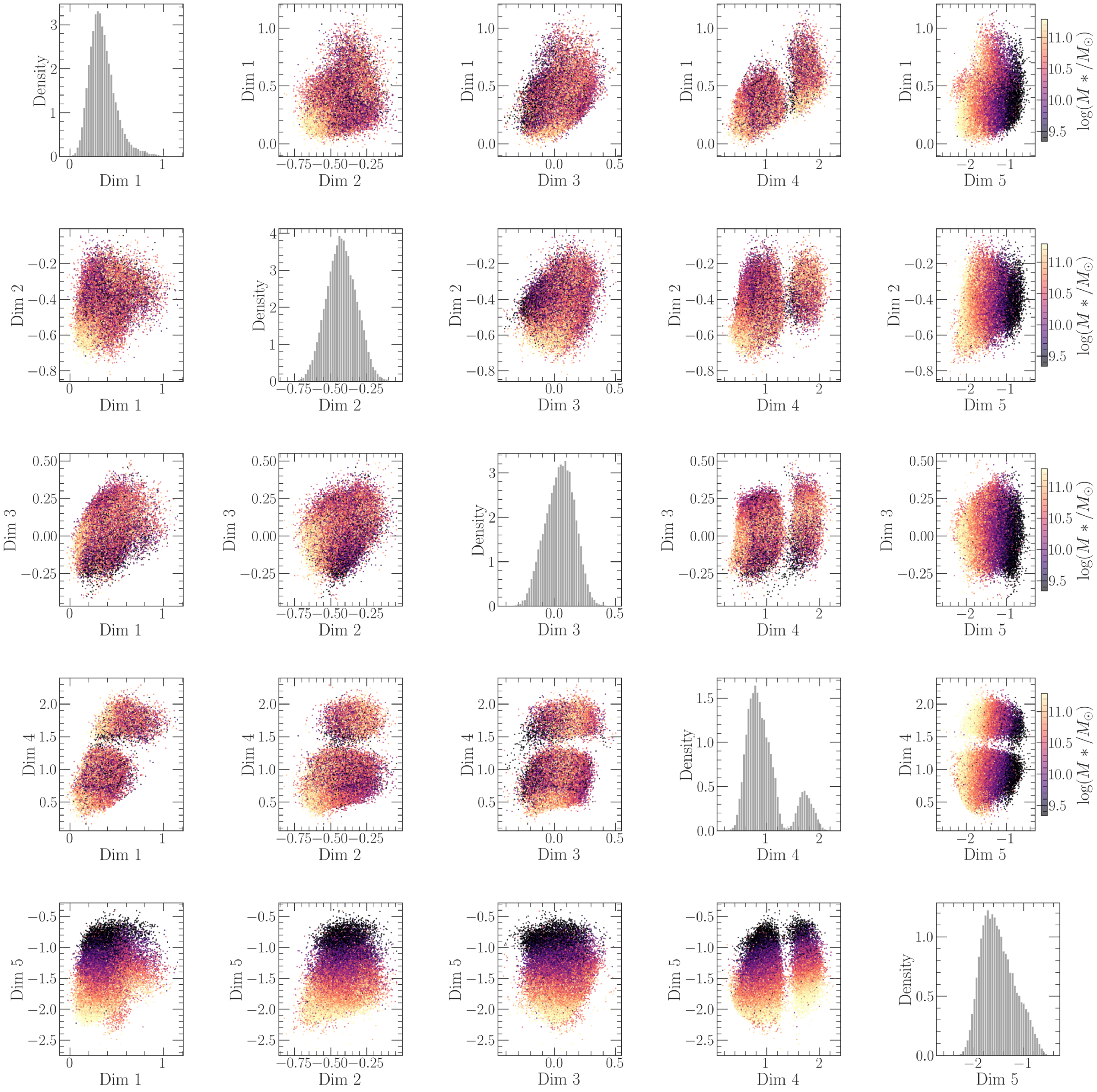}
    \caption{Same features as shown in Fig.~\ref{fig:corner_redshift}, but colored by the mean stellar mass of the galaxy derived in \citet{2022Zou_SED}}
    \label{fig:corner_mass}
\end{figure*}

\begin{figure*}
    \centering
    \includegraphics[width=\linewidth]{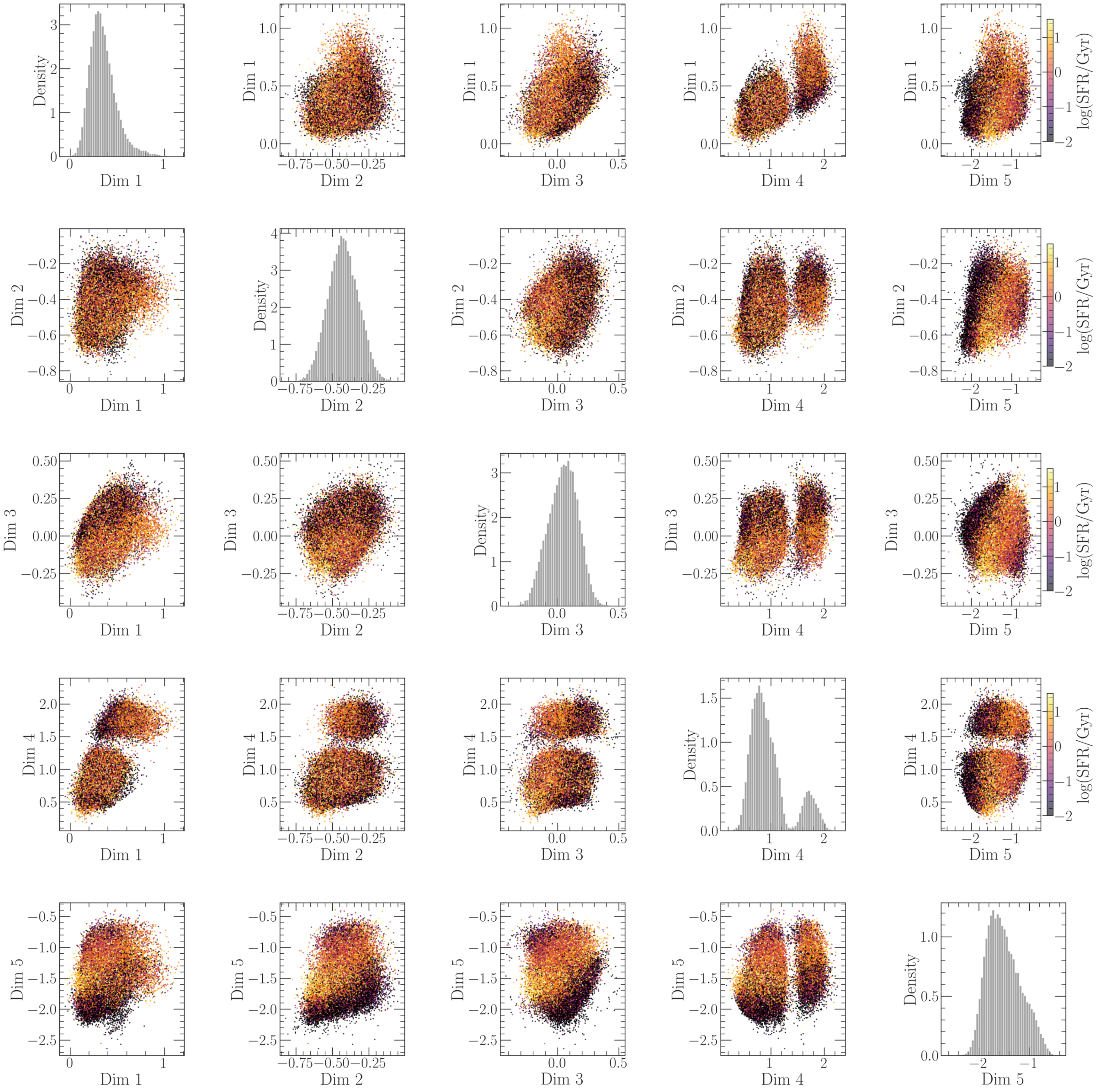}
    \caption{Same features as shown in Fig.~\ref{fig:corner_redshift}, but colored by the mean star-formation of the galaxy derived in \citet{2022Zou_SED}}
    \label{fig:corner_sfr}
\end{figure*}

The low dimensionality of our learned latent space also allows us to investigate its correlations with observed galaxy properties without the use of, e.g., UMAP. We color the latent features of validation-set galaxies by their SED-inferred properties reported in \citet{2022Zou_SED}. Our results for redshift, stellar mass, and star-formation rate are shown in Figure~\ref{fig:corner_redshift}, Figure~\ref{fig:corner_mass},  and Figure~\ref{fig:corner_sfr}, respectively. The redshift appears to be strongly correlated with all five latent dimensions, whereas the stellar mass and star-formation rate appear most correlated with the fifth latent dimension. In the fourth latent dimension, two clear clusters are observed. One of these clusters is associated with the lowest redshifts in the test set.

The correlations observed suggest that, despite the low model dimensionality, the learned latent features are predictive of SED-derived galaxy properties. We quantify this by training a neural spline flow to calculate the joint posterior of redshift, stellar mass, and star-formation rate of each galaxy conditioned on five latent features obtained from \model{}. We select 10 transforms and six hidden dimensions, each with 512 features. We train the model using 80\% of the galaxies from \model{}'s test set for 50 epochs and maximize the log-probability of the physical properties averaged across a batch size of 128 galaxies. 

\begin{figure*}[!ht]
    \centering
    \includegraphics[width=\linewidth]{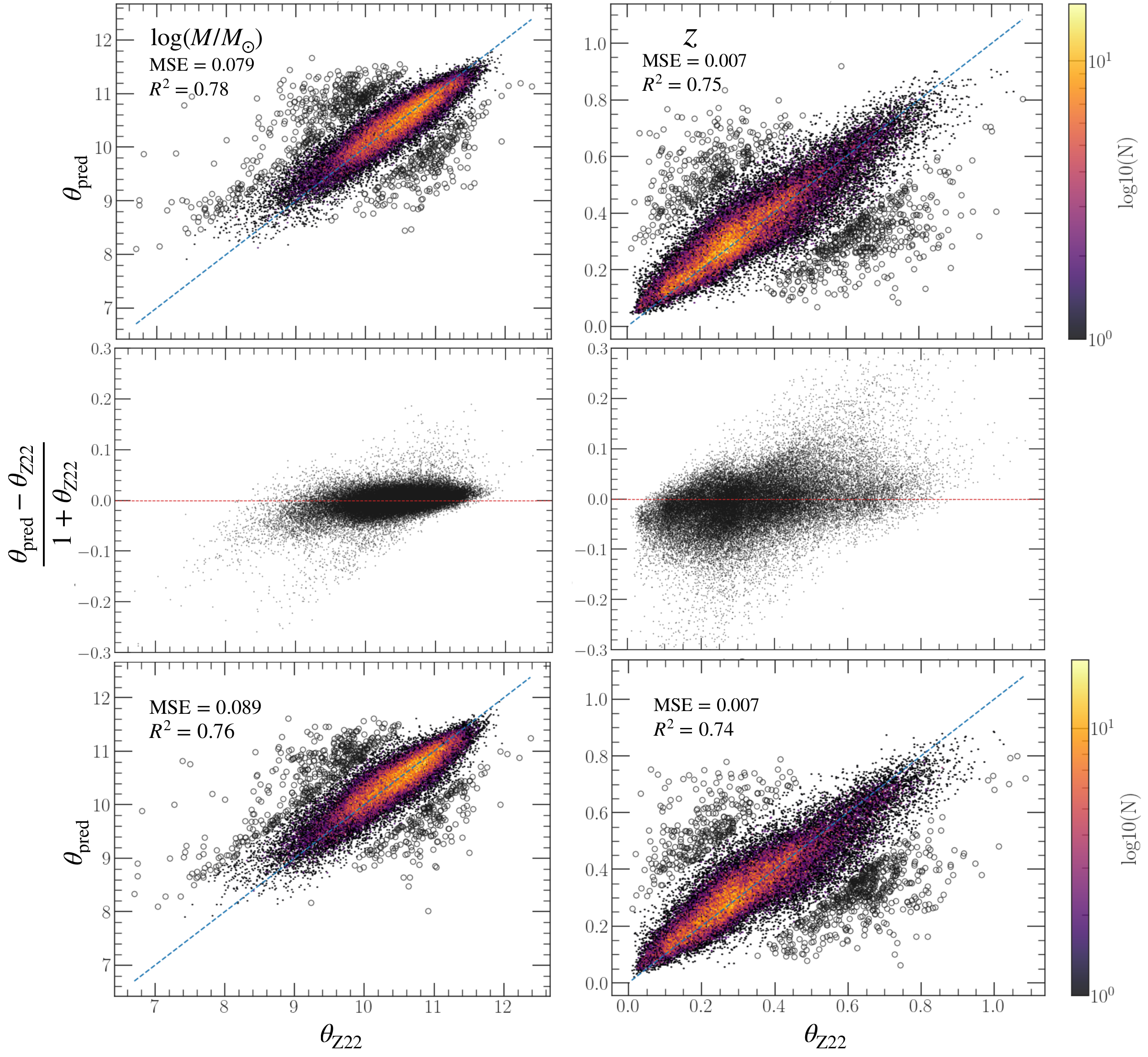}
    \caption{\textbf{Top Row:} Mean predictions from a joint conditional flow predicting galaxy stellar mass (left) and redshift (right) from the learned \model{} features of the validation set, compared to those inferred from SED fitting in \citet{2022Zou_SED}. Outliers $>3\sigma$ from the $y=x$ line are shown as open circles, and prediction mean-squared error (MSE) and coefficient of determination $R^2$ for each fit are shown top left. \textbf{Middle Row:} Fractional residuals for the predictions at top. \textbf{Bottom Row:} Performance for the same conditional flow trained on embeddings from \texttt{AION-1} \citep{2025Parker_AION}, a 300M-parameter foundation model for astrophysics.}
    \label{fig:flow_params}
\end{figure*}

Next, we evaluate the model on the held-out 20\% of galaxies. For each galaxy, we calculate the mean of 64 samples drawn for each property from the trained flow. We compare the predicted stellar mass and redshift against those reported by \citet{2022Zou_SED} in Figure~\ref{fig:flow_params} \footnote{We do not visualize the flow's performance on star-formation rate, as the \citet{2022Zou_SED} parameters were inferred from SED fits to optical photometry alone and the reported star-formation rates are less reliable (private communication). Nonetheless, we have found that the flow's performance on redshift and stellar mass improves with the inclusion of star-formation rate, likely for its ability to break degeneracies with the other two parameters.}.

The trained flow achieves strong performance in predicting physical parameters. We find a mean-squared-error (MSE) of $MSE=0.079$ and a coefficient of determination of $R^2=0.78$ for stellar mass, and a significantly lower $MSE=0.007$ but comparable $R^2=0.75$ for galaxy redshift. Our redshift $R^2$, obtained from photometry alone and with only five parameters, is only marginally lower than that reported in \citet{2024Parker_AstroCLIP} ($R^2=0.79$ for zero-shot and $R^2=0.78$ for few-shot), where the authors used contrastive learning to align galaxy images with spectra to extract physically-informative embeddings. We caution that their dataset is substantially smaller than the one considered here (their cross-matched spectroscopic and photometric sample contains 197,632 galaxies, versus our 6,050,000), and so the two results may not be directly comparable. Further, our comparison is to the galaxy properties estimated by \citet{2022Zou_SED} from extracted photometry, which may deviate from spectroscopic estimates. Our $R^2=0.78$ value for mass is higher than that obtained using galaxy images from \citet{2024Parker_AstroCLIP} for both the zero-shot and few-shot prediction task ($0.74$ and $0.73$, respectively); but is worse than their model performance using spectra ($0.87$ and $0.88$).

Next, we compare the performance of our trained conditional flow to an identical one trained on embeddings from \texttt{AION-1} \citep{2025Parker_AION}, a foundation model for astrophysics. \texttt{AION-1} is a transformer-based model trained by masked modeling of diverse astronomical data, which were tokenized with a series of modality-specific vector-quantized variational autoencoders. In addition to spectroscopy from Gaia \citep{2018Gaia}, DESI \citep{2016DESI_specsurvey}, and SDSS \citep{2000York_SDSS}, \textit{griz}-band DECaLS images for 122.7~M sources were used in training, allowing us to embed in-distribution imaging for our galaxies. We use the base model with 300~M total parameters.

Because \texttt{AION-1} was trained on $96\times 96\ \mathrm{px}$ images at the native resolution of DECaLS, we center crop our original \textit{grz}-band images to this size (instead of the $72\times 72\ \mathrm{px}$ representation used by \model{})\footnote{We also trained a model using the preprocessed $72\times 72\ \mathrm{px}$ images resampled to $96\times 96\ \mathrm{px}$ using \texttt{Pytorch}'s bilinear interpolation routine, but found lower $R^2$ and higher MSE values for both redshift and stellar mass}. We directly embed these images and mean pool over the 576 spatial tokens as recommended in the associated documentation, arriving at a 768-dimensional embedding for each galaxy. We then train a conditional flow to predict the redshift, stellar mass, and star-formation rate of each galaxy, training for 500 epochs and saving the performance of the model with lowest loss on the validation set. 

We plot the model's mean predictions for redshift and stellar mass in the bottom row of Figure~\ref{fig:flow_params}. The AION-conditioned flow achieves nearly identical performance to that conditioned on \model{} embeddings ($R^2=0.76$ for stellar mass, compared to $R^2=0.78$ for \model{}), suggesting that the significantly larger latent space is \textit{not} substantially more predictive of these inferred properties than those produced by our model. We doubled the number of hidden features in our conditional flow (from 512 to 1024) to accommodate a more complex mapping between \texttt{AION} latents and physical properties, but find no difference in performance. 

Finally, we evaluate the calibration of our conditional flow using the Tests
of Accuracy with Random Point (TARP) Expected Coverage Probability package \citep{2023Lemos_TARP}. The procedure selects a random parameter vector, builds a series of concentric spherical credible sets around it, and records how often the true parameter falls inside each set. 

\begin{figure}
    \centering
    \includegraphics[width=\linewidth]{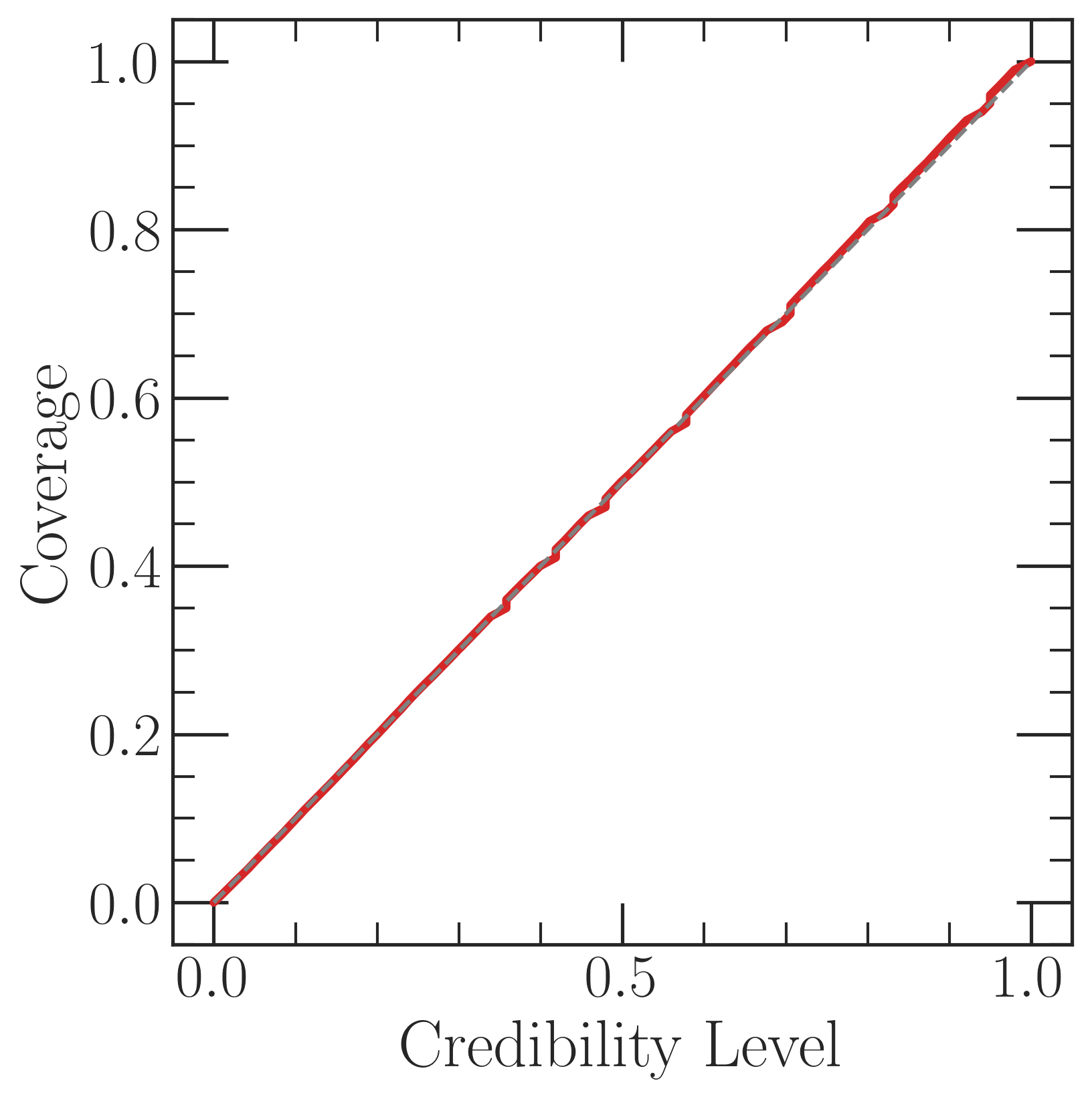}
    \caption{Coverage of the conditional flow predicting redshift, stellar mass, and star-formation rate of the central galaxy, showing unbiased posteriors for the validation set.}
    \label{fig:param_tarp}
\end{figure}

We present the results calculated for 100 bins in $\alpha\; \in \;[0,1]$ across the test set of our conditional flow in Figure~\ref{fig:param_tarp}. In the limit of perfect calibration, the confidence level matches the coverage probability (black dashed line). We find no strong bias in the reported posteriors of the test set. 

\subsection{Inference of Morphological Properties}\label{subsec:morphological_properties}

Next, we investigate the correlation between the learned latent features and the labeled morphological properties of each galaxy from Galaxy Zoo. 

We download the detailed morphological measurements from \citet{2022Walmsley_GalaxyZoo}\footnote{\url{https://zenodo.org/records/4573248}}, which consolidates 7.5 million individual labels for 314,000 galaxies within the DECaLS data release. Galaxies were classified according to a hierarchical decision tree, with broad morphological properties at the top level of the tree (``featured'' or ``smooth'') and granular features at the bottom level (such as the presence of rings or dust lanes). 

We cross-match our full DECaLS dataset with the Galaxy Zoo catalog, and find 33,667 total galaxies within the \citet{2022Zou_SED}, DECaLS, and \citet{2022Walmsley_GalaxyZoo} catalogues. Because of \model{}'s compact latent space, we limit our classification to labels requiring two or fewer dependent questions. We retrieve the vote fractions for four binary labels, with the positive labels defined as ``edge-on disk'', ``having spiral arms'', ``merger'' (combining galaxies labeled with \texttt{merging\_major-disturbance\_fraction} or \texttt{merging\_minor-disturbance\_fraction}), and ``round''. We include only galaxies with confident labels, and impose a minimum voting fraction of 50\% to consider a galaxy as having either a positive or a negative label (we lower this fraction to 20\% for the merging classifier to increase the number of examples in the dataset). 

\begin{figure*}
    \centering
    \includegraphics[width=\linewidth]{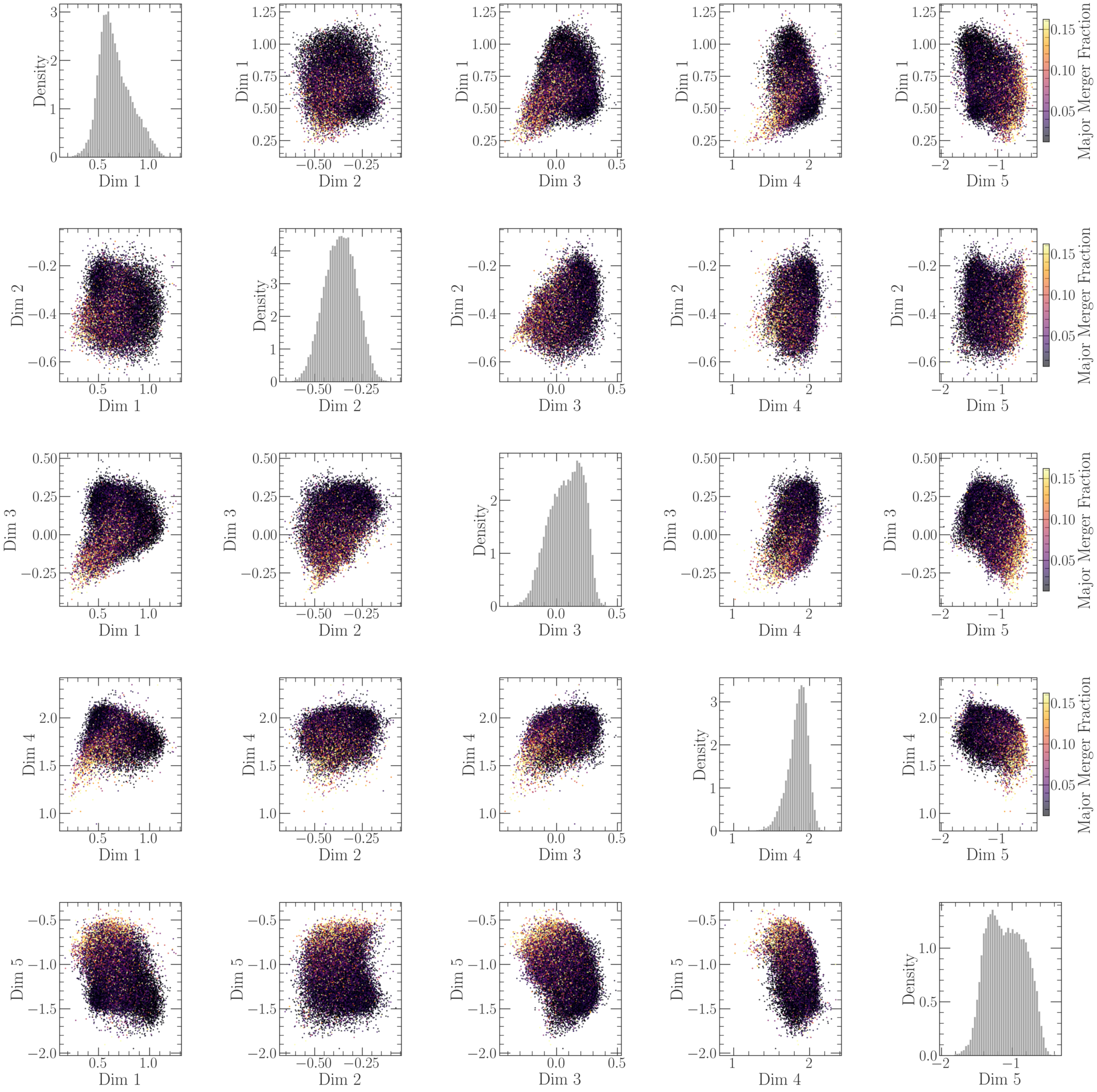}
    \caption{Corner plot of all latent features for the sample of galaxies cross-matched with the Galaxy Zoo catalog from \citet{2022Walmsley_GalaxyZoo}. Features are colored by the fraction of Galaxy Zoo volunteers that labeled the galaxy as a major merger (see text for details).}
    \label{fig:corner_merging}
\end{figure*}

\begin{figure*}
    \centering
    \includegraphics[width=\linewidth]{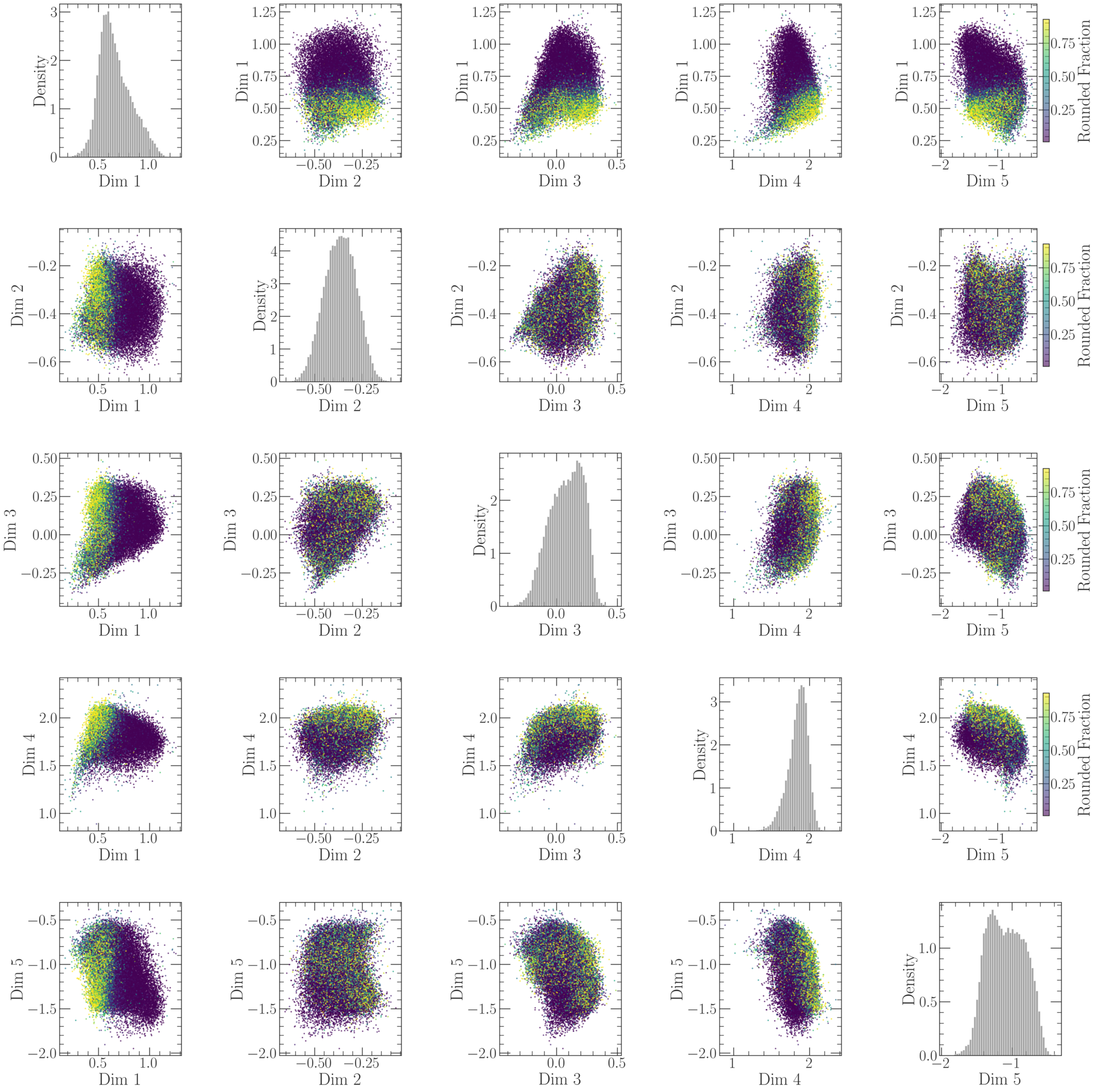}
    \caption{Same features as shown in Fig.~\ref{fig:corner_merging}, but colored by the fraction of volunteers identifying a galaxy as round.}
    \label{fig:corner_round}
\end{figure*}

\begin{figure*}
    \centering
    \includegraphics[width=\linewidth]{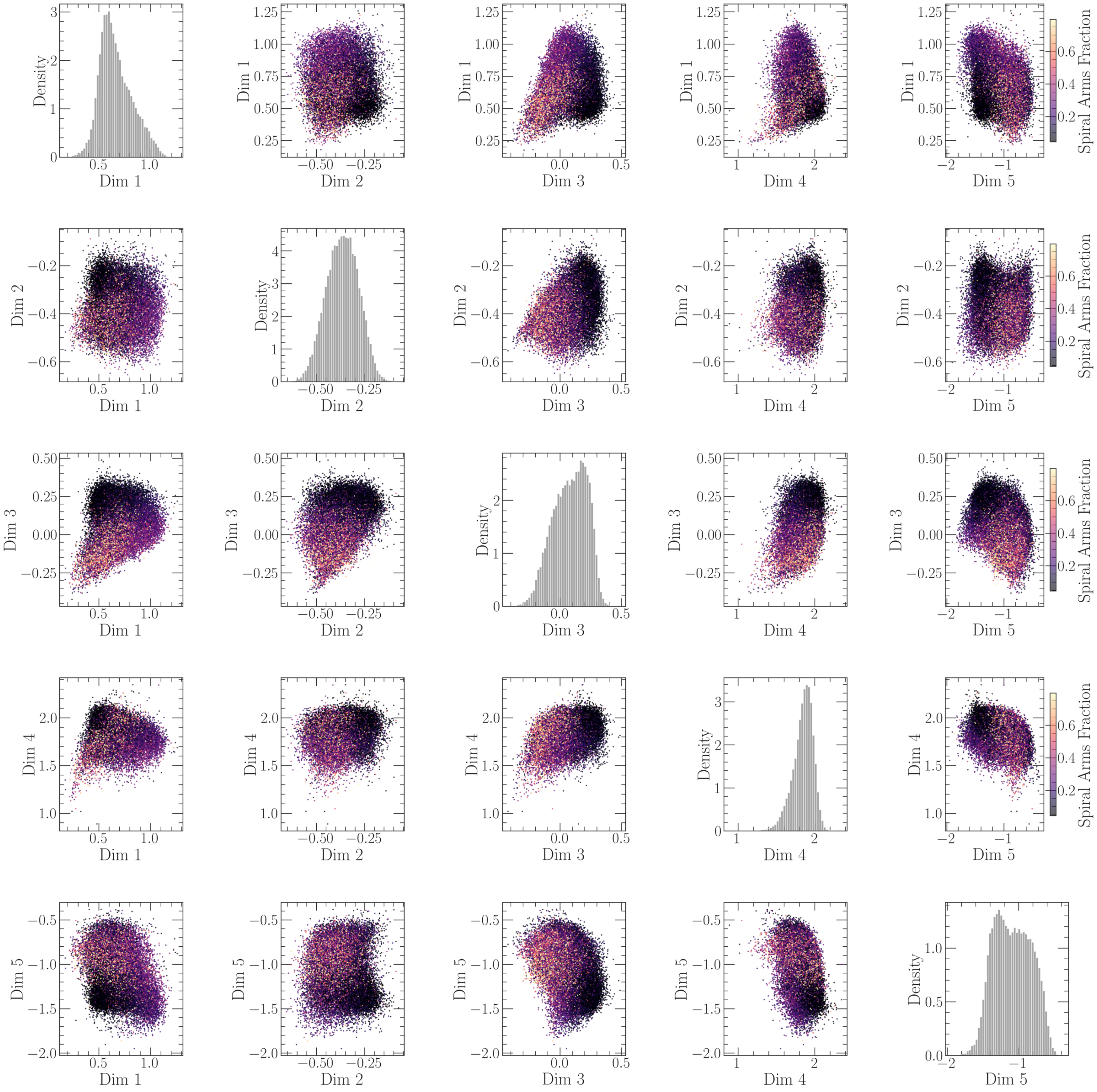}
    \caption{Same features as shown in Fig.~\ref{fig:corner_merging}, but colored by the fraction of volunteers identifying a galaxy as having spiral arms.}
    \label{fig:corner_spiralarms}
\end{figure*}

We visualize the distribution of Galaxy Zoo labels in the latent space in Figures~\ref{fig:corner_merging}, \ref{fig:corner_round}, and \ref{fig:corner_spiralarms}. As with physical properties, galaxies with similar morphological labels are clustered in the latent space; major mergers are visually distinguished by latent features 4 and 5, and round galaxies have low values of latent feature 1.

Next, we train a series of simple Multi-Layer Perceptron (MLP) heads to classify key morphological properties of each galaxy. Each MLP consists of 4 residual and 4 linear layers, each with 512 hidden neurons, a GELU activation function \citep{2016Hendrycks_GELU}, and a dropout fraction of 0.1. We train for 200 epochs using the standard \verb|Adam| optimizer \citep{2014AdamOptimizer} and a learning rate of $10^{-3}$. 

To account for the large imbalance in positively labeled features, we adopt a weighted random sampler in training with weights corresponding to the inverse of the number of instances of each positive label across the full dataset. We calculate the performance of our classifier in a 5-fold cross-validation split of the cross-matched data, each time training on 80\% of the data and testing on the remaining 20\%.

\begin{figure*}[!ht]
    \centering
    \includegraphics[width=\linewidth]{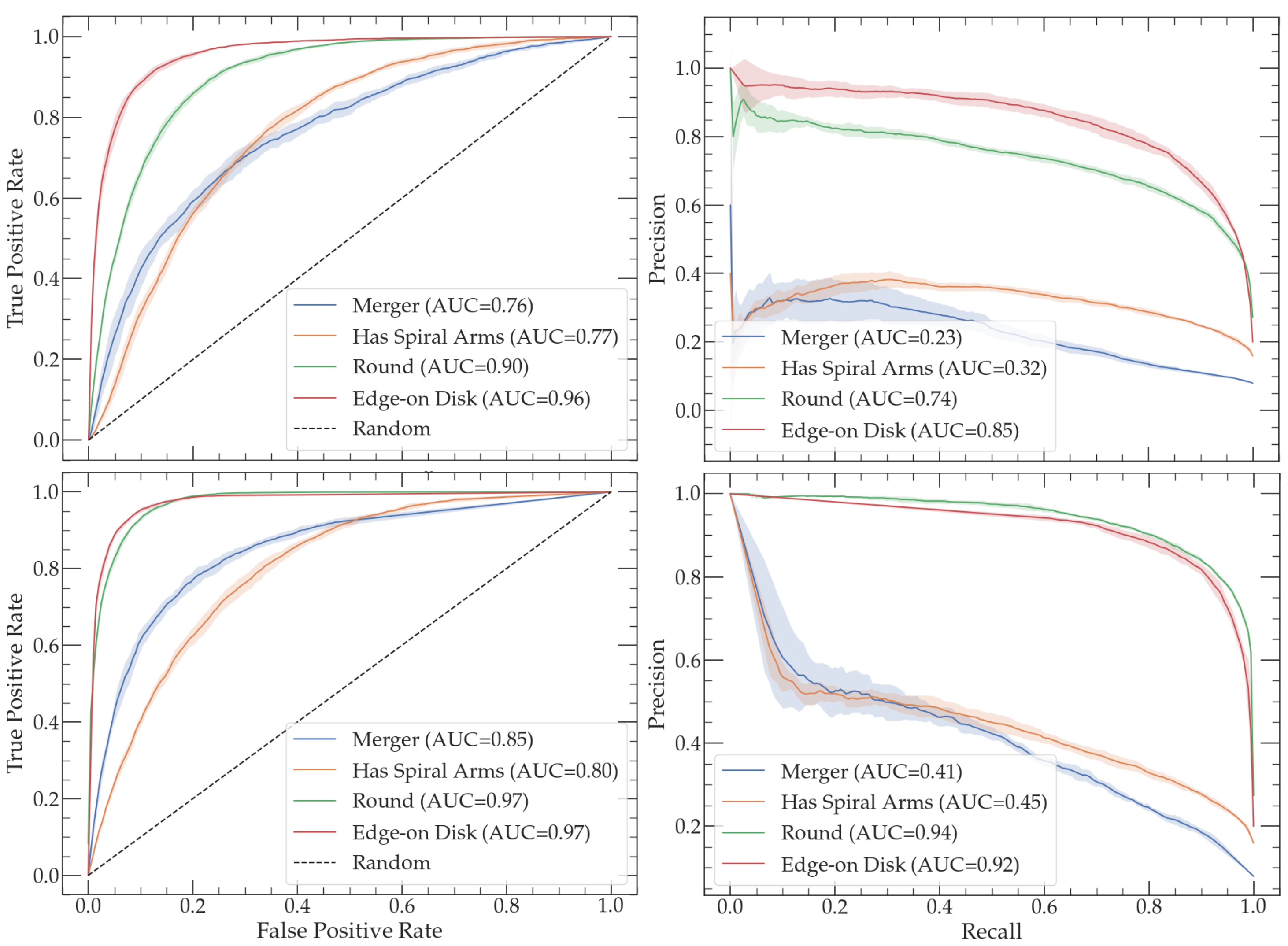}
    \caption{\textbf{Top Row:} Receiver Operator Characteristic curve (left) and Precision-Recall curve (right) associated with the binary morphological classification tasks described in \textsection\ref{subsec:morphological_properties}. Mean AUC values are given in the legend and shaded regions correspond to the 1$\sigma$ spread across 5 cross-validation splits. \textbf{Bottom Row:} The same plots for binary classifiers trained on embeddings from \texttt{AION-1} \citep{2025Parker_AION}, a general-purpose multi-modal model for astrophysics.}
    \label{fig:roc_curve}
\end{figure*}

\begin{figure*}
        \centering
    \includegraphics[width=0.8\linewidth]{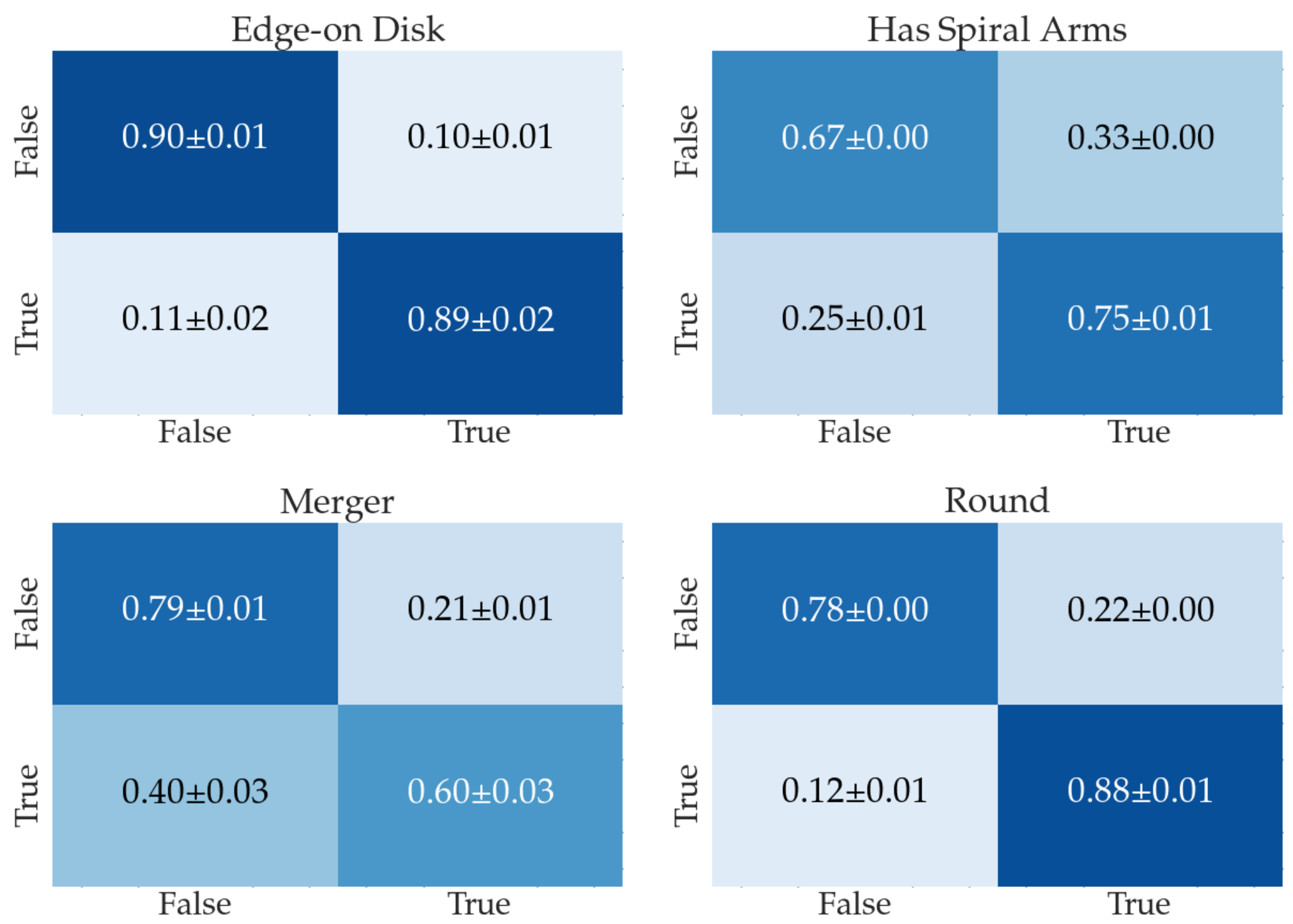}
    \caption{Completeness confusion matrices for the same classification tasks as in Fig.~\ref{fig:roc_curve}. The means and 1$\sigma$ uncertainties across 5 cross-validation splits are listed for each classifier. Performance is worst for classifying galaxy mergers, likely due to the reliance on complex morphological data not well-captured by our compact latent space.}
    \label{fig:morph_cm}
\end{figure*}

We present a Receiver Operator Characteristic (ROC) curve and a Precision-Recall (PR) curve for each binary classifier in Figure~\ref{fig:roc_curve}, and the confusion matrices associated with each model and normalized by number of true instances of each class (showing completeness/recall) in Figure~\ref{fig:morph_cm}. We also calculate the area under the ROC curve (AUROC) and the area under the PR curve (AUPRC). Both approach unity in the limit of perfect classification; a classifier performing no better than random will have AUROC$=$0.5 and AUPRC equal the fraction of positive labels in the test set (0.08, 0.16, 0.27, and 0.20 for ``merger'', ``spiral arms'', ``round'', and ``edge-on disk'' labels, respectively).

The classifier for edge-on disk galaxies achieves strongest performance, with an average AUROC$=$0.96 and AUPRC$=$0.85. \model{}'s latent features are similarly informative of galaxy roundness: our classifier achieves mean values of AUROC$=$0.90 and AUPRC$=$0.74. The presence of spiral arms, a label requiring two dependent questions in the original schema, is classified less accurately but higher than random chance (AUROC$=$0.77, AUPRC$=$0.32). 

The worst-performing classifier, and that with the largest variance in performance across five data folds, is the merger classifier. We calculate a mean AUROC$=$0.76 and AUPRC$=$0.23. We attribute this in part to the small number of merging galaxies for training either \model{} in the first stage or the binary classifier in the second (1,094 galaxies are listed with \texttt{merging\_major-disturbance\_fraction}$>$0.2, just over 3\% of the full cross-matched sample). \model{}'s compact latent space and the segmentation mask  used in training may also partially be to blame, as galaxies with irregular off-center structure may not be associated with the central galaxy and subsequently masked.

As with the conditional flow experiments in \textsection\ref{subsec:physical_properties}, we now compare our results to a series of identical binary classifiers trained on \texttt{AION} embeddings. We present results in the bottom row of Figure~\ref{fig:roc_curve}. In contrast with the inference of physical properties, here we observe a statistically-significant improvement in performance across all labels: the mean AUPRC value nearly doubles from $0.23$ for \model{}-predicted mergers to $0.41$ for \texttt{AION}-predicted mergers, and from $0.32$ to $0.45$ in predicting spiral arms. We propose potential reasons for this in the Discussion. Nonetheless, we still find that these two labels are predicted with the least precision and the highest variance across folds compared to predicting the 'round' and 'edge-on disk' labels, likely due to the ambiguity of the classifications and the limited number of positive examples for training.

\subsection{Similarity Searches}\label{subsec:similarity_search}

Finally, we use the learned latent space to perform k-nearest neighbor similarity search for galaxies in the test set. For an embedded representation a galaxy $z_i$, we retrieve the nearest neighbors $z_j$ with lowest distance in embedding space. We compare two distance measures: the Euclidean distance, defined as $d_E(z_i,z_j)=||z_i-z_j||_2$; and the Mahalanobis distance, which additionally considers the covariance matrix $\Sigma$ between latent features: $d_M(z_i,z_j)=\sqrt{(z_i - z_j)^T \Sigma^{-1} (z_i - z_j)}$. We do not compute the cosine similarity (the normalized scalar product) between latent features; though this has been done to retrieve neighbors in other works (e.g., \citealt{2024Parker_AstroCLIP}), the low dimensionality of the latent space and the large relative scale of dimensions 4 and 5 renders many objects with similar vector directionality in this space.

\begin{figure*}[h]
    \centering
    \includegraphics[width=\linewidth]{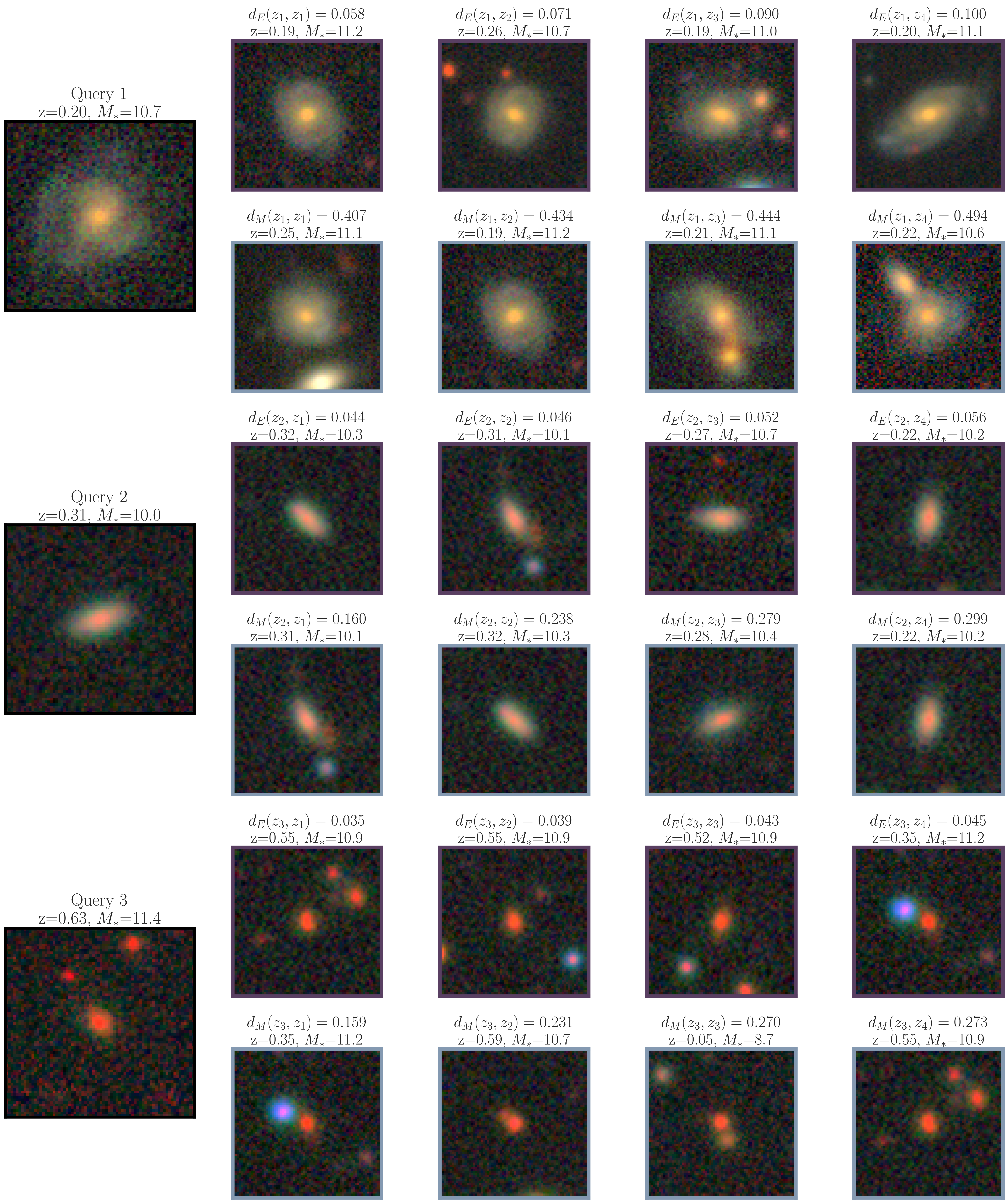}
    \caption{Three randomly selected galaxies (left column) and the four neighbors with lowest Euclidean distance (top row) and Mahalanobis distance (bottom row) in the learned latent space. The distance to each neighbor, and the redshift and stellar mass of each galaxy from \citet{2022Zou_SED}, is given at top.}
    \label{fig:neighbors}
\end{figure*}

We randomly select three galaxies from our test set and retrieve four test-set galaxies with lowest $d_E,d_M$ distances. We present the retrieved galaxies in Figure~\ref{fig:neighbors}, along with their photometric redshift and stellar mass estimates from \citet{2022Zou_SED}. The visual and physical similarities between queried and retrieved galaxies, despite differences in background sources, indicates that the latent features of \model{} can be useful for comparative studies of galaxy evolution and targeted searches of anomalous galaxy populations.

\section{Discussion}\label{sec:discussion}
We have presented \model{}, a representation learning model with a diffusion-based reconstruction objective trained on multi-band imaging from the DECaLS survey. We probe the information content of the five learned latent features through a series of downstream tasks, including the inference of spectroscopic properties and morphological features. The model's strong performance suggests that the architecture presented in \cite{2025Shen_Perceiver} is able to extract highly compact, semantically-meaningful information using a fully unsupervised objective. 

The ability to rapidly infer physical parameters competitive with state-of-the-art SED models using only three filtered galaxy images is encouraging. Encoding galaxy images and obtaining posteriors from the conditional flow takes substantially less time than the $\sim$10 hrs required for rigorous SED fitting using modern approaches \citep{2021Johnson_Prospector}, and orders of magnitude faster than existing simulation-based inference models optimized for this task \citep{2023Wang_SBI}. Our approach makes it feasible with minimal compute resources to obtain physical properties for all 20B galaxies imaged by the Vera C. Rubin Observatory across ten years. Neural network-based feature extraction techniques like the one presented here will help bring scalability and interpretability to petabyte-scale data exploration among astronomical surveys, and facilitate the search for new galaxy sub-types.

\model{}'s compact latent space has allowed us to visualize the correlation between latent features and catalog-level information about each galaxy. While the latent dimensions are not factorized, we find preliminary evidence that the image information most strongly correlated with physical properties are concentrated to a subset of the five latent dimensions. As a result, \model{}'s extracted features strike a balance between interpretability, which may surpass that of larger models such as \citet{2024Parker_AstroCLIP}; and feature expressivity, which surpasses that of properties typically reported in astronomical catalogs. In future work, we plan to use \model{} to embed the images of transient host galaxies and include the latent features in a suite of models for time-domain classification and anomaly detection (with, e.g., \texttt{reLAISS}; \citealt{2025Reynolds_reLAISS}).

The learned latent space is invariant to galaxy orientation (Fig.~\ref{fig:ddim_samples}), likely as a consequence of the limited capacity of the five latent tokens. While this reduces information content that may be valuable for targeted downstream applications, it benefits tasks where orientation is a nuisance parameter (e.g., redshift and mass estimation).

Despite promising performance, we have also found that \model{} features may not be strongly predictive of irregular morphological features such as evidence of galaxy mergers. The segmentation mask used to identify pixels within an image's central region of interest may present an overly-rigid design for highly segmented or irregularly-shaped galaxies at low redshift. An extension of this framework could be to impose a continuous-weighted mask (taking the form of, e.g., a Gaussian distribution) centered on the galaxy image.  Alternative techniques, such as a luminosity-weighted mask or a modification to the attention weights associated with each image patch in the encoder, may result in superior representations without increasing the latent dimensionality. However, given the similarly reduced performance of the \texttt{AION}-trained classifiers on these labels compared to predicting round galaxies and edge-on disks, larger labeled training sets will likely bring additional performance gains.

Finally, we compared the predictive power of our galaxy embeddings for physical and morphological quantities to those produced by the recently released foundation model \texttt{AION}. We observe substantial improvements for all morphological labels but no gain in predicting redshift or stellar mass (from \citealt{2022Zou_SED}), which is unexpected given \texttt{AION}’s much larger latent dimensionality (576 versus 5 for \model{}). Several factors could explain this. First, fine-grained morphological features (e.g., features predictive of spiral structure or mergers) may benefit from a larger model and a higher-dimensional latent space, whereas global properties with substantial pixel-level redundancy (such as redshift and stellar mass) may be sufficiently captured by a much smaller representation. Second, while we evaluated a larger conditional flow for \texttt{AION} embeddings, we did not perform an exhaustive hyperparameter sweep, which could yield improved performance for the baseline. Differences in architecture and training objectives also challenge our ability to direct interpret the performance results. Finally, most labels in \citep{2022Zou_SED} are derived from SED fits to extracted photometry; large spectroscopic samples obtained by upcoming galaxy surveys (e.g., DESI; \citealt{2016DESI_specsurvey}) will provide tighter physical constraints and may reveal more informative discrepancies between embedding techniques.

\section{Acknowledgments}
We thank Carolina Cuesta-Lazaro and John Wu for conversations that improved this manuscript. This work is supported by the National Science Foundation under Cooperative Agreement PHY-2019786 (The NSF AI Institute for Artificial Intelligence and Fundamental Interactions, http://iaifi.org/). 

The Villar Astro Time Lab acknowledges support through the David and Lucile Packard Foundation, National Science Foundation under AST-2433718, AST-2407922 and AST-2406110, as well as an Aramont Fellowship for Emerging Science Research. 

The computations in this paper were run on the FASRC Cannon cluster, supported by the FAS Division of Science Research Computing Group at Harvard University.

The Legacy Surveys consist of three individual and complementary projects: the Dark Energy Camera Legacy Survey (DECaLS; Proposal ID \#2014B-0404; PIs: David Schlegel and Arjun Dey), the Beijing-Arizona Sky Survey (BASS; NOAO Prop. ID \#2015A-0801; PIs: Zhou Xu and Xiaohui Fan), and the Mayall z-band Legacy Survey (MzLS; Prop. ID \#2016A-0453; PI: Arjun Dey). DECaLS, BASS and MzLS together include data obtained, respectively, at the Blanco telescope, Cerro Tololo Inter-American Observatory, NSF’s NOIRLab; the Bok telescope, Steward Observatory, University of Arizona; and the Mayall telescope, Kitt Peak National Observatory, NOIRLab. Pipeline processing and analyses of the data were supported by NOIRLab and the Lawrence Berkeley National Laboratory (LBNL). The Legacy Surveys project is honored to be permitted to conduct astronomical research on Iolkam Du’ag (Kitt Peak), a mountain with particular significance to the Tohono O’odham Nation.

NOIRLab is operated by the Association of Universities for Research in Astronomy (AURA) under a cooperative agreement with the National Science Foundation. LBNL is managed by the Regents of the University of California under contract to the U.S. Department of Energy.

This project used data obtained with the Dark Energy Camera (DECam), which was constructed by the Dark Energy Survey (DES) collaboration. Funding for the DES Projects has been provided by the U.S. Department of Energy, the U.S. National Science Foundation, the Ministry of Science and Education of Spain, the Science and Technology Facilities Council of the United Kingdom, the Higher Education Funding Council for England, the National Center for Supercomputing Applications at the University of Illinois at Urbana-Champaign, the Kavli Institute of Cosmological Physics at the University of Chicago, Center for Cosmology and Astro-Particle Physics at the Ohio State University, the Mitchell Institute for Fundamental Physics and Astronomy at Texas A\&M University, Financiadora de Estudos e Projetos, Fundacao Carlos Chagas Filho de Amparo, Financiadora de Estudos e Projetos, Fundacao Carlos Chagas Filho de Amparo a Pesquisa do Estado do Rio de Janeiro, Conselho Nacional de Desenvolvimento Cientifico e Tecnologico and the Ministerio da Ciencia, Tecnologia e Inovacao, the Deutsche Forschungsgemeinschaft and the Collaborating Institutions in the Dark Energy Survey. The Collaborating Institutions are Argonne National Laboratory, the University of California at Santa Cruz, the University of Cambridge, Centro de Investigaciones Energeticas, Medioambientales y Tecnologicas-Madrid, the University of Chicago, University College London, the DES-Brazil Consortium, the University of Edinburgh, the Eidgenossische Technische Hochschule (ETH) Zurich, Fermi National Accelerator Laboratory, the University of Illinois at Urbana-Champaign, the Institut de Ciencies de l’Espai (IEEC/CSIC), the Institut de Fisica d’Altes Energies, Lawrence Berkeley National Laboratory, the Ludwig Maximilians Universitat Munchen and the associated Excellence Cluster Universe, the University of Michigan, NSF’s NOIRLab, the University of Nottingham, the Ohio State University, the University of Pennsylvania, the University of Portsmouth, SLAC National Accelerator Laboratory, Stanford University, the University of Sussex, and Texas A\&M University.

BASS is a key project of the Telescope Access Program (TAP), which has been funded by the National Astronomical Observatories of China, the Chinese Academy of Sciences (the Strategic Priority Research Program “The Emergence of Cosmological Structures” Grant \#XDB09000000), and the Special Fund for Astronomy from the Ministry of Finance. The BASS is also supported by the External Cooperation Program of Chinese Academy of Sciences (Grant \#114A11KYSB20160057), and Chinese National Natural Science Foundation (Grant \#12120101003, \#11433005).

The Legacy Survey team makes use of data products from the Near-Earth Object Wide-field Infrared Survey Explorer (NEOWISE), which is a project of the Jet Propulsion Laboratory/California Institute of Technology. NEOWISE is funded by the National Aeronautics and Space Administration.

The Legacy Surveys imaging of the DESI footprint is supported by the Director, Office of Science, Office of High Energy Physics of the U.S. Department of Energy under Contract No. DE-AC02-05CH1123, by the National Energy Research Scientific Computing Center, a DOE Office of Science User Facility under the same contract; and by the U.S. National Science Foundation, Division of Astronomical Sciences under Contract No. AST-0950945 to NOAO.

GAMA is a joint European-Australasian project based around a spectroscopic campaign using the Anglo-Australian Telescope. The GAMA input catalog is based on data taken from the Sloan Digital Sky Survey and the UKIRT Infrared Deep Sky Survey. Complementary imaging of the GAMA regions is being obtained by a number of independent survey programmes including GALEX MIS, VST KiDS, VISTA VIKING, WISE, Herschel-ATLAS, GMRT and ASKAP providing UV to radio coverage. GAMA is funded by the STFC (UK), the ARC (Australia), the AAO, and the participating institutions. The GAMA website is http://www.gama-survey.org/.

\section*{References}

\medskip
\raggedbottom
\bibliography{references}

\end{document}